\documentclass{ws-procs975x65}

\begin{document}

\title{\uppercase{Loop Quantization of Shape Dynamics}}
\author{\uppercase{Tim A. Koslowski}}
\address{Department of Mathematics and Statistics, University of New Brunswick\\ Fredericton, NB, E3B 5A3, Canada\\ email: \texttt{t.a.koslowski@gmail.com}}

\bodymatter

\begin{abstract}
  Loop Quantum Gravity (LQG) is a promising approach to quantum gravity, in particular because it is based on a rigorous quantization of the kinematics of gravity. A difficult and still open problem in the LQG program is the construction of the physical Hilbert space for pure quantum gravity. This is due to the complicated nature of the Hamilton constraints. The Shape Dynamics description of General Relativity (GR) replaces the Hamilton constraints with spatial Weyl constraints, so the problem of finding the physical Hilbert space reduces to the problem of quantizing the Weyl constraints. Unfortunately, it turns out that a loop quantization of Weyl constraints is far from trivial despite their intuitive physical interpretation. A tentative quantization proposal and interpretation proposal is given in this contribution. 
\end{abstract}
Shape Dynamics (SD) is a canonical formulation of GR that is locally indistinguishable from the ADM formulation. The manifest equivalence can be seen when the ADM system is evolved in constant mean extrinsic curvature (CMC) gauge and a special gauge is used to evolve SD: It turns out that the initial value problem and the equations of motion of the two theories coincide in this case. SD was constructed in \cite{SD}. On a compact Cauchy surface $\Sigma$ without boundary, one can express SD in extrinsic curvature variables as a system with first class constraints, which generate internal rotations, diffeomorphisms and conformal transformations:
\begin{equation}
 \begin{array}{c}
 G(\lambda)=\int_\Sigma d^3x\,\lambda^i{\epsilon_{ij}}^kK^j_aE^a_k,\,\,\,H(v)=\int_\Sigma d^3x K^i_a(\mathcal L_v E)^a_i,\\
 C(\rho)=\int_\Sigma d^3x \rho (K^i_aE^a_i-\frac 2 3 \tau \sqrt{|E|}).
 \end{array}
\end{equation}
Time evolution in York time $\tau$ is generated by the conformal York Hamiltonian, which is the on-shell volume of $\Sigma$
\begin{equation}
 H_{SD}=\int_\Sigma d^3x \sqrt{|E|} \Omega_o^6[K,E]
\end{equation}
where metric variables are $g_{ab}=|E|E^i_aE^j_b\delta_{ij}$ and $\pi^{ab}=E^a_kE^d_k/(2|E|)(K^i_c\delta^b_d-K^i_d\delta^b_c)E^c_i$, when the rotation constraints hold. The functional $\Omega_o[K,E]$ is defined as the solution of the Lichnerowicz--York equation
\begin{equation}
  8\Delta \Omega_o=R[E]\Omega_o+\frac 3 8 \tau^2 \Omega_o^5-\pi^{ab}[K,E](g_{ac}[E]g_{bd}[E]-\frac 1 3 g_{ab}[E]g_{cd}[E])\pi^{cd}[K,E]/\sqrt{|E|} 
\end{equation}
These variables satisfy canonical Poisson brackets.\\
{\bf{Ashtekar variables.}} I follow Thiemann's convention \cite{Thiemann} and introduce Ashtekar variables through a canonical transformation generated by the generating functional $F=\int_\Sigma d^3x\left(\beta K^i_a\tilde{E}^a_i+\tilde E^a_i\Gamma^i_a(\tilde{E})\right)$, where tildes denote transformed variables and where $\Gamma^i_a$ denotes the components of the spin connection expressed as a functional of the densitized inverse triad, leading to the canonical pair of Ashtekar variables
\begin{equation}
 \tilde A_a^i=\beta K^i_a+\Gamma^i_a(\tilde E),\,\,\,\tilde E^i_a=E^i_a/\beta,
\end{equation}
where $\beta$ denotes the Immirzi parameter. After integration by parts, one finds a simple form for the diffeomorphism and Gauss constraints
\begin{equation}
 G(\lambda)=\int_\Sigma d^3x \lambda^i \tilde D_a \tilde E^a_i,\,\,\,H(v)=\int_\Sigma d^3x \tilde A^i_a(\mathcal L_v E)^a_i,
\end{equation}
where $\tilde D_a \tilde E^a_i=\epsilon_{ijk} \tilde A^j_a\tilde E^a_k+\tilde E^a_{i,a}$. The conformal constraints take the unpleasant form
\begin{equation}\label{equ:confConstr}
 C(\rho)=\int_\Sigma d^3x \rho \left((\tilde A^i_a-\Gamma^i_a(\tilde E))\tilde E^a_i-\frac {2\,\tau}{3\, \beta^{3/2}}\sqrt{|\tilde E|}\right),
\end{equation}
which depends nonlinearly on the momenta $\tilde E^a_i$ and on their derivatives. After some standard simplifications, one finds that the Hamiltonian takes essentially the same form as in extrinsic curvature variables:
\begin{equation}\label{equ:Hamiltonian}
 H_{SD}=\beta^{-3/2} \int_\Sigma d^3x \sqrt{|\tilde E|} \Omega^6_o[K(\tilde A,\tilde E),\beta \tilde E]. 
\end{equation}
The nonlinear dependence of $C(\rho)$ on the momenta $\tilde E$ already indicates that a conformal transformation of $\tilde A_a^i$ can not be expressed in terms of $\tilde A_a^i$ alone; in fact a conformal transformation takes $\tilde A_a^i \to e^{-4\phi}\tilde A_a^i+(1-e^{-4\phi})\Gamma^i_a(\tilde E)+2\epsilon^{ijk}(\tilde E^{-1})^j_a \tilde E^b_k \phi_{,b}$. Hence, infinitesimal conformal transformations can not be quantized as vector fields on the space of Ashtekar connections. 
\\
{\bf{Kinematic Loop Quantization:}} A description of the loop quantization procedure goes beyond the scope of this contribution. I refer the reader to textbooks\cite{Thiemann} and a forthcoming publication that explains detail behind this short contribution. I will only recall some relevant facts about the kinematics of LQG:\\
1. A basis for the kinematic Hilbert space of LQG is given by diffeomorphism classes of spin-network functions of Ashtekar parallel transports.\\
2. The standard geometric operators of LQG yield the following interpretation of spin networks: The spin quantum numbers on edges of the spin network functions are quanta of area of transversally intersecting surfaces. The intertwiners at vertices represent quanta of volume for a region containing the vertex as well as quantized dihedral angles between surfaces that intersect at the vertex.\\
3. Spin network functions are eigenfunctions of spatial quantum geometry in the sense that a large enough set of commuting area, volume and angle quantum numbers completely determine the spin network function.\\
4. It is important for the following to note the fact that angle operators do not commute and that the general commutator of two angle operators depends on the volume operator.\\
{\bf{Conformal Invariance:}} Since a quantization of the conformal constraints can not be achieved as vector fields on the space of Ashtekar connections, one needs to adopt a different quantization strategy. One particular strategy would be to consider classical conformal invariants and to implement the quotient by conformal transformations through building equivalence classes of spin network functions that yield the same conformal invariants. \\
The classical part is simple: Angles and the vanishing of area and volume are conformal invariants, while area and volume are ``pure gauge'' in the sense that any nonvanishing area/volume can be transformed into any other nonvanishing area/volume. One would thus naively expect that one can implement the aforementioned quantization strategy by removing all area and volume quantum numbers from the spin network functions while retaining the angle quantum numbers. This simple idea is obstructed by the fact that the algebra generated by angle and volume operators is noncommutative, so the operators can not be simultaneously diagonalized.\\
One way to solve this problem is to simultaneously diagonalize the volume operators and area operators (corresponding to surfaces that do not intersect a vertex) only. One then finds a basis labeled by elements $ |[\gamma];j_1,...,j_n;V_1,...,V_m;\textrm{more}\rangle$, where $[\gamma]$ denotes the equivalence class of graphs $\gamma$ that are mapped into each other by a diffeomorphism, $j_i$ denotes the spin labels on edges and $V_i$ the volume quantum numbers at vertices and ``more'' denotes the quantum numbers at vertices that are not subsumed by the volume quantum numbers\footnote{For the sake of brevity, I have neglected the issue of graph automorphisms, which complicate the solution of the diffeomorphism constraint and imply relations among the quantum numbers $j_1,...,j_n;V_1,...,V_m;\textrm{more}$.}. One can then implement the quantization strategy through constructing equivalence classes
\begin{equation}
 |[\gamma];j_1,...,j_n;V_1,...,V_m;\textrm{more}\rangle \sim |[\gamma];k_1,...,k_n;U_1,...,U_m;\textrm{more}\rangle.
\end{equation}
The physical Hilbert space of loop quantized SD is then supposed to be spanned by these equivalence classes. Unfortunately, this destroys the interpretation of the data contained in ``more'' in terms of eigenvalues of the angle operators of LQG.\\
{\bf{Open Problems:}} There are two main open problems with the loop quantization of SD: (1) Finding a physical interpretation of the vertex labels and (2) constructing a physically acceptable dynamics, i.e. a ``quantization'' of equ. (\ref{equ:Hamiltonian}). One can follow many ideas to attack these two problems, but I believe that a sensible answer can only be found in a renormalization group (RG) setting: (1) Macroscopic angles can only be expected to arise in a coarse-grained setting. The effective ``renormalized'' angle operators should commute and their relation with the fundamental data ``more'' is expected to be nontrivial. (2) In the RG setting, one would also not try to quantize equ. (\ref{equ:Hamiltonian}), but rather study whether any simple bare Hamiltonian leads to an approximation of equ. (\ref{equ:Hamiltonian}) in a physically sensible coarse grained limit. 
\\
{\bf{Acknowledgements:}} This work was supported by the Natural Science and Engineering Research Council of Canada.

\end{document}